\documentclass[conference]{IEEEtran}
\IEEEoverridecommandlockouts
\usepackage{cite}
\usepackage{amsmath,amssymb,amsfonts}
\usepackage{algpseudocode}
\usepackage{algorithm}  
\usepackage{graphicx}
\usepackage{textcomp}
\usepackage{tabularx}
\usepackage{xcolor}
\usepackage{subcaption}
\usepackage{booktabs} 
\usepackage{makecell}
\usepackage{multirow}
\def\BibTeX{{\rm B\kern-.05em{\sc i\kern-.025em b}\kern-.08em
    T\kern-.1667em\lower.7ex\hbox{E}\kern-.125emX}}

\begin{document}

\title{A Deep and Transfer Learning Approach for Handover Management in O-RAN}

\author{
\IEEEauthorblockN{Ioannis Panitsas, Akrit Mudvari, Ali Maatouk, Leandros Tassiulas}
\IEEEauthorblockA{
Department of Electrical and Computer Engineering, Yale University, USA}
}

\maketitle

\begin{abstract}

Despite advancements in the Open-Radio Access Network (O-RAN) architecture and its applications, handover management remains an open problem without a comprehensive methodology for orchestrating seamless cell transitions within the O-RAN context. Existing predictive handover algorithms have not been designed inherently with scalability in mind, which is crucial for large-scale cellular networks where a Radio Intelligent Controller (RIC) can manage multiple distributed cells. In addition, their static model architectures limit adaptability to dynamic RAN topologies, making them irresponsive and impractical for real-world deployments. To address these challenges, we propose an O-RAN-compliant algorithm with an adaptive and flexible design that can be implemented as an xApp in the Near-Real-Time RIC to facilitate high-accuracy predictive handover decisions. We demonstrate that our algorithm outperforms state-of-the-art predictive handover methods, achieving over 95\% classification accuracy in large-scale networks while reducing the number of handovers by 46\% compared to the standard 3GPP handover algorithm. Moreover, it adapts effectively to dynamic RAN topologies with a varying number of base stations, requiring at least 30\% less retraining time while maintaining high accuracy.

\end{abstract}


\section{Introduction}
Next-generation (NextG) cellular networks are evolving into disaggregated, programmable, and open systems to support emerging vertical use cases and enable new services \cite{b1}. For example, the Open Radio Access Network (O-RAN) architecture \cite{b16}, proposed and standardized by the O-RAN Alliance, embraces disaggregation and openness to reduce both operational and capital expenditures by fostering competition. It also promotes multi-vendor interoperability and enables the integration of Artificial Intelligence and Machine Learning (AI/ML) capabilities into the RAN for enhanced network optimization \cite{b2}, \cite{b3}.

The O-RAN Alliance is continually proposing and refining use cases to promote this open architecture \cite{b5}. One use case that has been identified is the \textit{handover management}, which remains a critical area of academic focus. Handover management is not a new network related problem; it has been studied since the days of 2G networks and involves transferring an active connection from one Base Station (BS) to another to maintain seamless service continuity. Despite extensive research, handover optimization in NextG networks remains an open problem \cite{b1}, \cite{b2}, \cite{b5} due to new and unexplored challenges \cite{b3}, such as: 1) network densification and access heterogeneity; 2) utilization of new frequency bands (e.g., mmWave and THz); 3) emergence of ultra-reliable, low-latency services with minimal interruption time; 4) dynamic, high-speed user mobility; and 5) integration of non-terrestrial and terrestrial networks all of which further complicate handover decision-making.

In recent years, significant research efforts have focused on data-driven approaches, specifically applying ML techniques in handover optimization due to their powerful ability to accurately model the handover decision process and adapt to complex, dynamic, and temporal network conditions \cite{b4}. Predictive handover management has gained a lot of attention [6-10], due to its potential to enhance user experience by minimizing service interruptions during mobility. For instance, in \cite{b6} and \cite{b7}, the authors proposed Deep Neural Network (DNN) architectures to predict radio link failures and early handovers. In \cite{b8}, the authors introduced Long Short-Term Memory (LSTM) based architectures to predict future channel quality reference signals. Similarly, in \cite{b9}, a Recurrent Neural Network (RNN) architecture is presented to learn the optimal timing and destination BS for initiating handovers. Finally, in \cite{b11}, the authors formulated a joint connection management and load balancing problem and explored the use of reinforcement learning to approximate the solution.

Despite the contributions of these works, there is still progress to be made. Although the O-RAN architecture has matured since its introduction in 2018, a comprehensive methodology for handover management within the O-RAN context still remains absent. While the work in \cite{b11} addresses several issues in that direction, it lacks details on data collection, model training, and deployment within the O-RAN context. Second, while many predictive handover algorithms exist in the literature \cite{b6}, \cite{b7},\cite{b8},\cite{b9},\cite{b10} none of them have been evaluated in scenarios where scalability is a real concern, neglecting the fact that, in production cellular networks, a RAN Intelligent Controller (RIC) can control multiple distributed cells. Finally, a significant gap remains in evaluating and testing these approaches in scenarios where dynamic events can occur, potentially impacting both the RAN topology and the algorithm architectures in real time. For instance, in NextG networks, Unmanned Aerial Vehicles (UAV) mounted BSs are expected to be deployed during short-term traffic surges, such as social events, to boost network capacity \cite{b12}.  As a result, these algorithms must be inherently designed with flexible architectures capable of adapting to dynamic network environments without complete retraining, as frequent retraining with each topology change is computationally inefficient and impractical for production cellular networks.

\subsection{Methodology and Contributions}
To address these challenges, we propose a predictive handover algorithm with an adaptable architecture consisting of three components: 1) Encoder: a flexible DNN with a dynamic input size that scales linearly with the number of active BSs connected to the RIC; 2) Stacked LSTM Component: a multi-layer LSTM model designed to extract and learn complex temporal patterns; and 3) Decoder: a configurable DNN with a dynamic output size that generates a probability distribution across cells, enabling the selection of the optimal BS for handover in the near future.  The architecture operates on temporal channel quality measurements reported by the User Equipment (UE) and supports the integration of additional dynamic features if needed. We formulate the handover decision process as a multi-class classification problem, with the objective of minimizing the cross-entropy loss between the predicted and ground-truth BS. Our results indicate that our algorithm achieves over 95\% classification accuracy in large-scale networks and significantly reduces the number of handovers compared to traditional approaches. Moreover, it adapts efficiently to dynamic RAN topologies, requiring at least 30\% less retraining time while maintaining high accuracy. Overall, the contributions of this paper are the following: 

\noindent 1) \textit{Methodology:} We present a comprehensive methodology outlining the implementation steps of a predictive handover algorithm within the O-RAN context.

\noindent 2) \textit{Prediction Algorithm:} We propose a learning-based algorithm compliant with O-RAN specifications for predictive handover management. Simulation results show that our approach outperforms four state-of-the-art handover prediction algorithms in large-scale RAN scenarios.

\noindent 3) \textit{Architecture:} We design a novel and adaptable architecture that adjusts to a dynamic number of base stations, reducing retraining time by leveraging transfer learning techniques to reuse learned weights in the updated model while maintaining high classification performance.

\section{Background}

\subsection{Brief Introduction to O-RAN}

O-RAN embraces the RAN disaggregation by splitting the monolithic BS into three functional units: the Central Unit (CU), Distributed Unit (DU), and Radio Unit (RU) \cite{b2}. The CU is responsible for functions such as radio resource control, connection management, and encryption while the DU manages real-time functions, including resource scheduling and error correction. The RU handles the transmission and reception of radio signals. The CU and DU, referred commonly as E2 nodes, are interconnected with software-based RICs via open interfaces. This enables telemetry data streaming from the E2 nodes and the implementation of closed loop control actions and policies using ML algorithms to optimize the RAN. There are two types of RICs that manage and control the RAN: the Near-Real-Time RIC, which operates on time scales between 10 milliseconds and 1 second and connects to the E2 nodes via the E2 interface, and the Non-Real-Time RIC, which operates on time scales greater than 1 second and can receive data from any functional unit of the BS through the O1 interface. ML models that are deployed in the Near-RT-RIC are called xApps, while those in the Non-RT RIC are called rApps. Each xApp can subscribe to specific RAN functions that are published from the E2 nodes to receive data that can be useful for predictions or to send control actions to the RAN. For instance, an xApp can utilize an E2 Service Model (E2SM) to receive Key Performance Metrics from the E2 nodes (E2SM-KPM) or to send a RAN Control action (E2SM-RC). Utilizing these service models, an xApp can actually control the RAN and make adaptive decisions based on real-time network conditions.

\subsection{Handover Procedure in 3GPP}

According to the 3GPP technical specification \cite{b15}, the handover procedure is typically triggered by measurement events (e.g., Event A3), where a UE detects that the Reference Signal Received Power (RSRP) from a neighboring BS exceeds that of its serving BS by at least a predefined threshold, known as the Handover Margin (HOM). If this condition persists for a specified duration, referred to as the Time To Trigger (TTT), the UE sends a measurement report to the serving BS. Upon receiving this report, the serving BS determines the optimal target BS to which the UE should be handed over, thereby completing the handover process. Despite its widespread adoption, this event-based procedure is highly sensitive to the selection of the triggering parameters, HOM and TTT. Low values may lead to excessive or unnecessary handovers (e.g., ping-pong effects), while high values can delay handover decisions, increasing the risk of radio link failures. These limitations motivate the development of a more adaptive and predictive handover strategy, which we outline in the following section.

\begin{figure*}[htbp]
    \centering
    \includegraphics[width=0.65\textwidth]{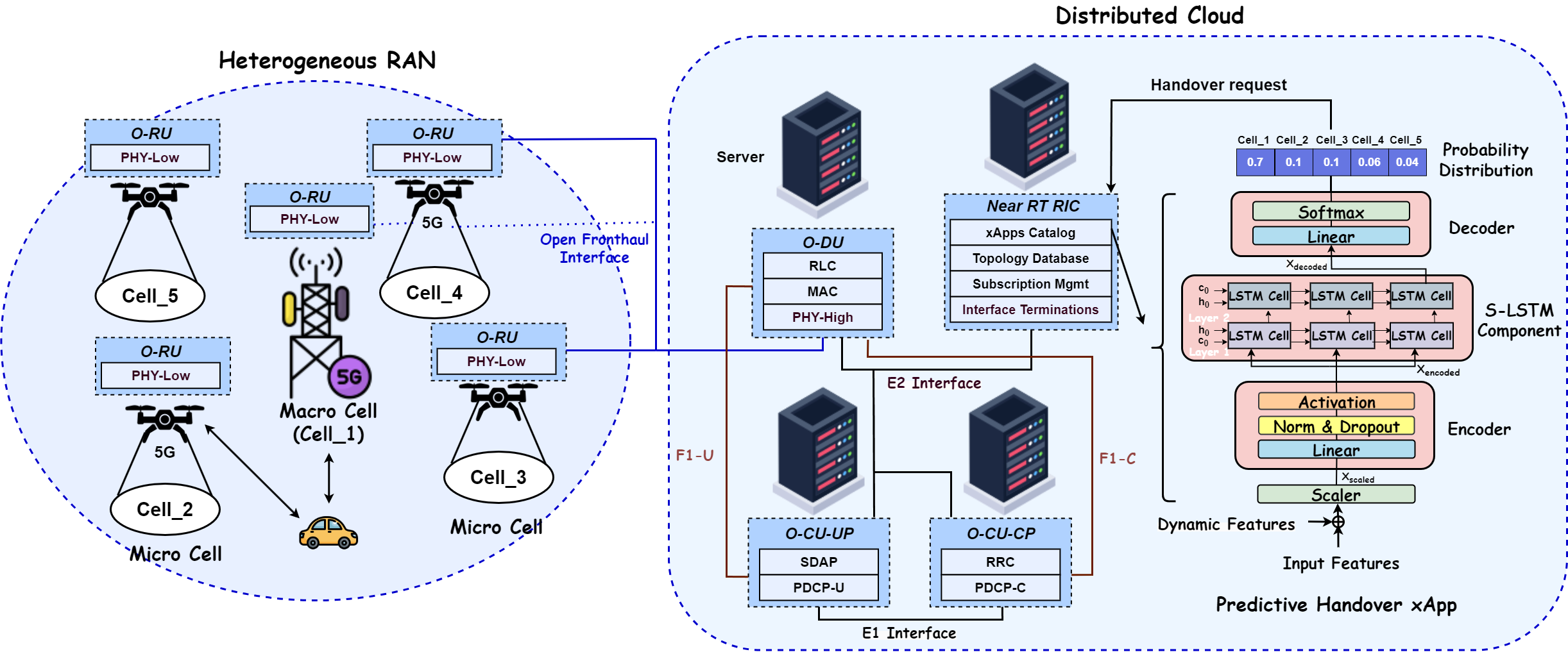}  
    \caption{Our proposed predictive handover algorithm is implemented as an xApp in the near-RT RIC.}
    \label{fig:fullwidthfigure}
\end{figure*}

\section{System Design and Modeling}


\subsection{Problem Formulation and Design Considerations}
To avoid unnecessary handovers, often triggered by noisy channel measurements or suboptimal handover thresholds, we propose a UE-centric, learning-based handover algorithm that predicts the future serving BS of a UE based on recent temporal channel quality reports. Unlike traditional handover mechanisms that rely on instantaneous signal thresholds or event-based triggers, our approach generates a stable BS association prediction over a future time window, filtering out unnecessary ping-pong handovers. Our algorithm is O-RAN compliant and can be implemented as an xApp in the Near-RT RIC, utilizing two E2SMs: E2SM-KPM, for collecting UE-reported channel measurements, and E2SM-RC, for dispatching control actions to the serving and target BSs involved in the handover process.

Building on the previous, we formalize the BS-UE association problem as a supervised learning task, where the objective is to predict the target BS that a UE will attach to in the near future, based on past temporal channel measurements and BS associations. More specifically, let $\mathbf{X} \in \mathbb{R}^{K \times N}$ represent a matrix of temporal UE channel quality measurements, where $K$ is the number of past measurements, and $N$ is the number of BSs managed by the Near-RT RIC for mobility control within a given tracking area. Each element $x_{k,n} \in \mathbf{X}$ denotes the channel quality reported by the UE from BS $n \in N$ at time step $t - K + k$. In addition, let $\mathbf{Y} \in [0,1]^{W \times N}$ represent the ground truth BS associations of the UE over a window of $W$ time steps, encoded as a one-hot vector for a number of $N$ BSs. The objective of our algorithm is to produce a prediction matrix $\mathbf{C} \in [0,1]^{W \times N}$, where each row $\mathbf{C}_{w,:}$ encodes the predicted association scores for the $N$ candidate target BSs at time step $t + w$, such that the average cross-entropy loss between the ground truth labels $\mathbf{Y}$ and the predicted outputs $\mathbf{C}$ over the prediction window is minimized:
\begin{equation}
\mathcal{L} = -\frac{1}{W} \sum_{w=1}^{W} \sum_{n=1}^{N} \mathbf{Y}_{w,n} \log {\mathbf{C}}_{w,n}
\label{eq:cross_entropy_loss}
\end{equation}
Beyond accurate prediction, adaptability to evolving RAN topologies is a critical design consideration for our model. In operational environments, the number and spatial distribution of active BSs within a tracking area may change due to temporary deployments (e.g., UAV-based BSs for bursty traffic) or energy-saving deactivations. These variations directly impact both the input and output dimensions of the learning algorithm, as the set of candidate BSs managed by the Near-RT RIC evolves over time. To address this, we design our algorithm with a flexible model architecture capable of accommodating dynamic BS sets. Specifically, our approach supports variable input and output sizes and leverages transfer learning to reuse previously learned weights, thereby reducing retraining overhead. This adaptability ensures high predictive performance while enabling efficient and continuous operation under topological changes. We now detail the architecture of our learning algorithm and its components.

\subsection{Model Architecture} 

Our model architecture consists of three primary components: an Encoder, a Stacked Long Short-Term Memory (S-LSTM) module, and a Decoder, as illustrated in the right part of Fig.~1. The Encoder transforms the variable-length input matrix $\mathbf{X}$ into a fixed-size representation. It consists of a sequence of three fully connected Multi-Layer Perceptrons (MLPs), each with 128 neurons and ReLU activations. This transformation enables the model to flexibly handle variations in the number of BSs during dynamic events. The Encoder outputs a fixed-length embedding of size 128 that is passed to the recurrent backbone for sequential modeling. The S-LSTM module is composed of two stacked LSTM layers, each with a hidden size of 128, designed to capture temporal dependencies in the encoded sequence. This component is responsible for learning time-dependent patterns in the UE's channel quality time series. By introducing the Encoder before the S-LSTM layers, the model allows for lightweight adaptation: when topology changes occur, retraining the Encoder is significantly more efficient than retraining the entire recurrent structure. This design choice helps preserve the temporal learning capacity of the LSTM layers while enabling adaptation through Encoder retraining. Finally, the Decoder receives the final hidden state from the S-LSTM module and generates predictions for future UE–BS associations. It is implemented as a fully connected MLP, followed by a softmax activation to produce a probability distribution over the candidate target BSs. The UE is then associated with the BS corresponding to the maximum value in this distribution, representing the most likely serving BS. Note that the number of output neurons in the Decoder is dynamically adjusted to match the current number of active BSs in the network. A confidence threshold can also be applied to the maximum score to suppress uncertain predictions and ensure robust handover decisions before triggering control actions via the Near-RT RIC.

\subsection{Model Training}

Our model was trained offline using historical UE channel measurements stored in the internal database of the Near-RT RIC. To construct the training data, we formed input–label pairs by applying a sliding window of size $K$ over past channel quality reports, producing input matrices $\mathbf{X} \in \mathbb{R}^{K \times N}$. For each input window, the corresponding label was derived by identifying the most frequently associated BS over the subsequent $W$ time steps. This label was then replicated across the entire prediction window to encourage the learning of stable BS association patterns and minimize short-term fluctuations that could lead to unnecessary handovers. Each training sample was flattened and normalized before being passed through the model. Model training was supervised using the cross-entropy loss between the predicted associations and the ground truth labels. Stochastic gradient descent with a learning rate of 0.0001 was used to learn the model weights. Moreover, the structure of the model layers, along with other training hyperparameters, was selected based on a grid search to achieve optimal performance. To mitigate overfitting, batch normalization and dropout were applied during training. Finally, the complete training procedure is summarized in Algorithm~\ref{alg:model_training}.


    

\begin{algorithm}[t]
\small 
\caption{Model Training}\label{alg:model_training}
\textbf{Input:} 
Randomized parameters \(\theta\) of our model, learning rate \(\eta\), batch size \(B\), number of epochs \(E\), number of batches \(N_{\text{batches}}\), ground truth labels \(Y_{W, N}\) \\ 
\textbf{Output:} learned parameters \(\theta\) of the model
\begin{algorithmic}

\For{$e = 1$ \textbf{to} $E$}  
    \For{$b = 1$ \textbf{to} $N_{\text{batches}}$}  
        \State $\textnormal{X}_{\textnormal{scaled}} \gets \textnormal{Scaler}(\textnormal{X}_{K, N})$ 
        \State $\textnormal{X}_{\textnormal{encoded}} \gets \textnormal{Encoder}(\textnormal{X}_{\textnormal{scaled}})$ 
        \State $\textnormal{X}_{\textnormal{S-LSTM}} \gets \textnormal{S-LSTM}(\textnormal{X}_{\textnormal{encoded}})$ 
        \State $\textnormal{C}_{W, N} \gets \textnormal{Decoder}(\textnormal{X}_{\textnormal{S-LSTM}})$ 
        \State $\textnormal{L}(\theta) \gets -\frac{1}{B} \sum_{i=1}^{B}  \sum_{j=1}^{N} \textnormal{Y}_{i, j} \log(\textnormal{C}_{i, j}(\theta)) $ 
        \State \textnormal{via Back Propagation Through Time.} 
        \State \textnormal{Update} $\theta \leftarrow \theta - \eta \nabla_{\theta} \textnormal{L}$.
    \EndFor
\EndFor

\end{algorithmic}
\end{algorithm}


\section{Evaluation Results}

\begin{figure*}[t]
    \centering
    \begin{subfigure}{0.48\textwidth}
        \centering
        \includegraphics[width=\linewidth]{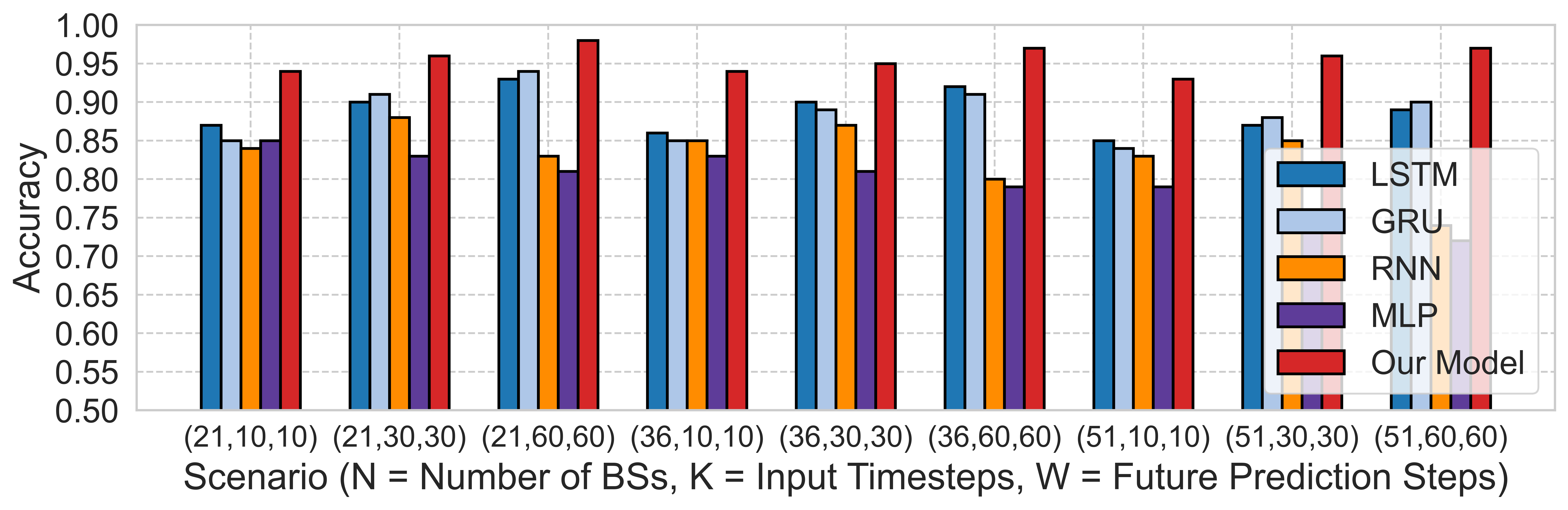}
        \caption*{(a) Accuracy}  
        \label{fig:accuracy}
    \end{subfigure}
    \hfill
    \begin{subfigure}{0.48\textwidth}
        \centering
        \includegraphics[width=\linewidth]{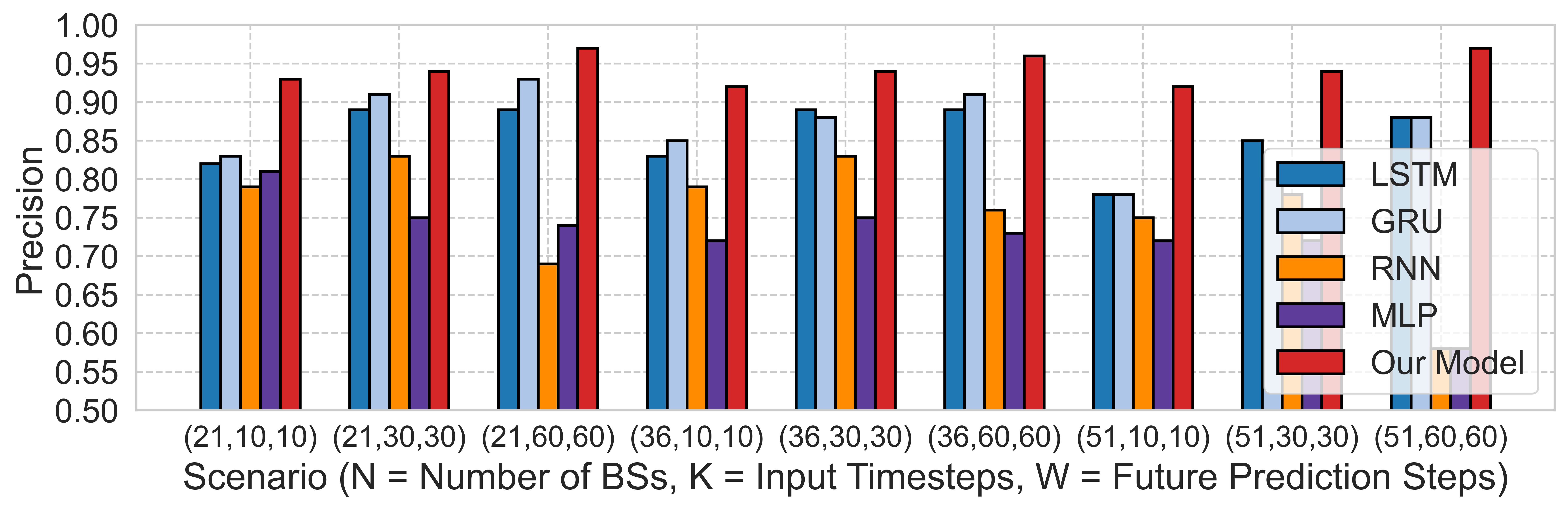}
        \caption*{(b) Precision}
        \label{fig:precision}
    \end{subfigure}

    \begin{subfigure}{0.48\textwidth}
        \centering
        \includegraphics[width=\linewidth]{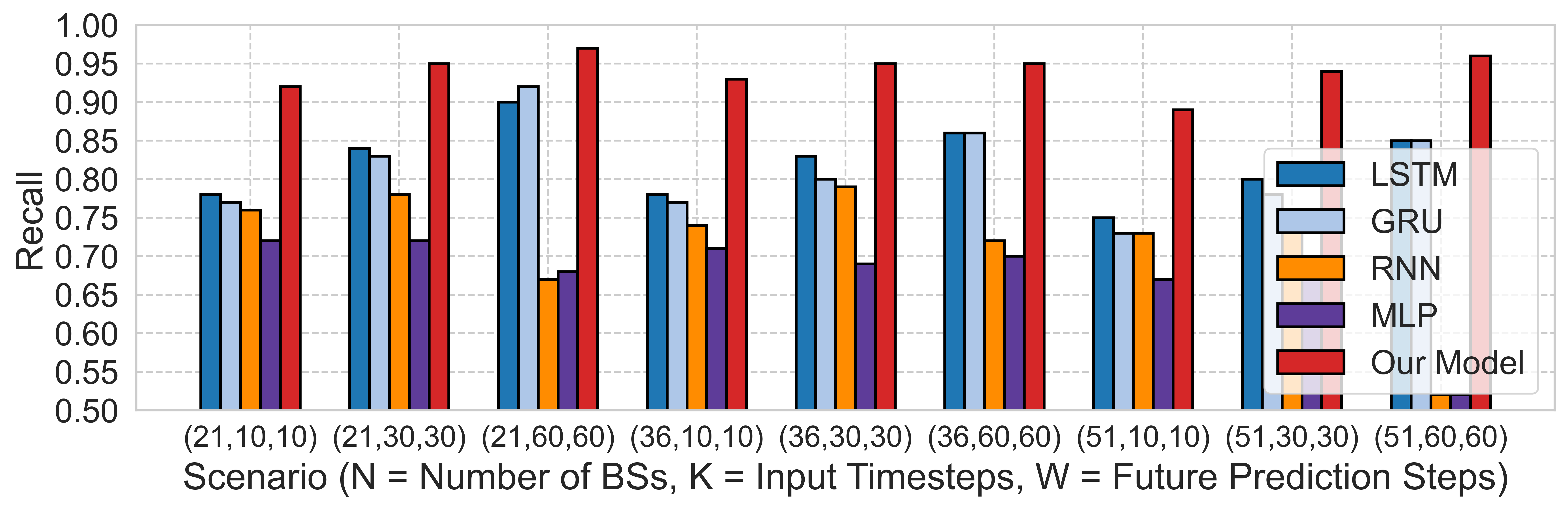}
        \caption*{(c) Recall}
        \label{fig:recall}
    \end{subfigure}
    \hfill
    \begin{subfigure}{0.48\textwidth}
        \centering
        \includegraphics[width=\linewidth]{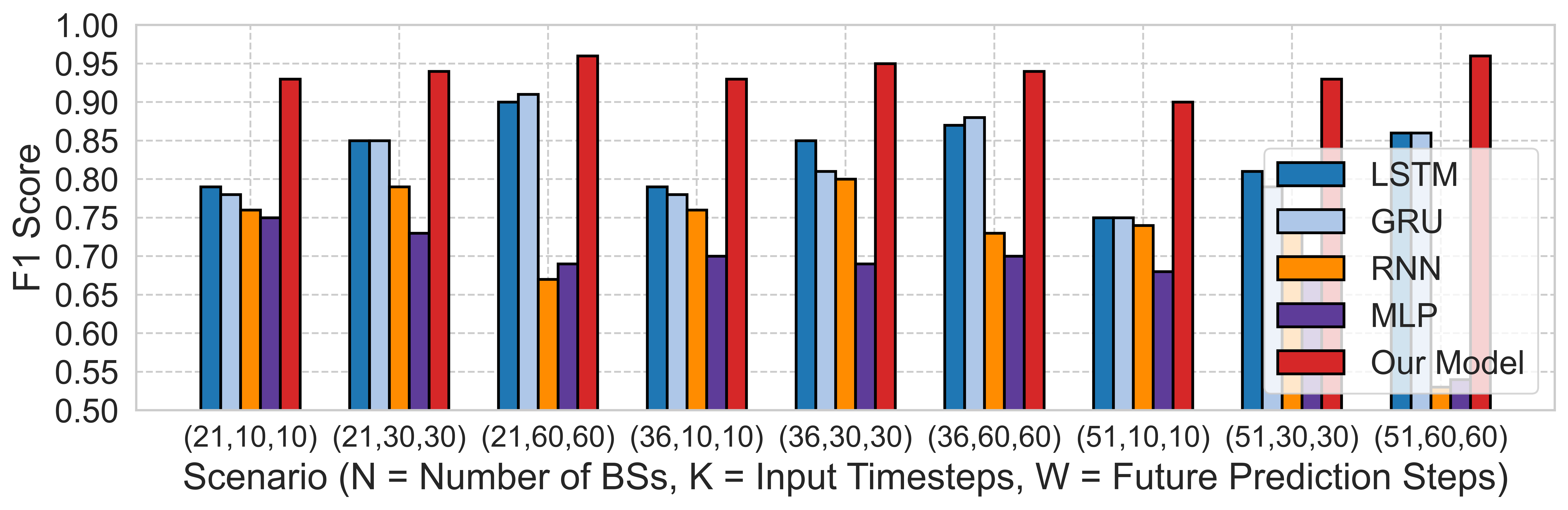}
        \caption*{(d) F1 Score}
        \label{fig:f1}
    \end{subfigure}

    \caption{Model performance across various network and model configurations $(N, K, W)$. The top left subplot (a) shows the accuracy, the top right (b) presents the precision, the bottom left (c) illustrates the recall, and the bottom right (d) displays the F1 Score for all models.}
    \label{fig:model-metrics}
\end{figure*}

\subsection{Network Environment}
In this work, we examine three simulated RAN environments composed of $N \in \{21, 36, 51\}$ BSs, which provide network access to end users.  We consider two types of cells: a macro cell that operates in the n5 band with a central frequency $f_1$ of 850 MHz, and a micro cell that operates in the n79 band with a central frequency $f_2$ of 4500 MHz. In this access network, a single macro cell provides wide-area coverage over an urban region, while the remaining $N-1$ micro cells are deployed within the same area to boost capacity in zones with high user density, as illustrated in the left part of Fig. 1. In this setup, we assume that the RU of the macro cell is deployed on a tall building at a height of 25 m, characterized by an Urban Macro (UMa) Non-Line-of-Sight (NLOS) path loss model ($ 13.54 + 39.08 \cdot \log_{10}(d) + 20 \cdot \log_{10}(f_1) $)\cite{b14}, where  $d$ is the distance between the RU and a UE. The macro BS operates with a transmission power of 40 dBm and uses an omnidirectional antenna with a gain of 3 dBi. On the other site, the RU of each micro cell is deployed on a UAV at a fixed location height of 80 m, and it is characterized by an Urban Micro (UMi) LOS path loss model ($32.4 + 21 \cdot \log_{10}(d) + 20 \cdot \log_{10}(f_2)
)$ \cite{b14}.
To simulate more realistic path losses due to signal degradation from environmental factors, we included shadow fading with a standard deviation of 7 dB for the macro cell and 4 dB for the micro cells. Each micro BS operates with a transmission power of 10 dBm, and uses an omnidirectional antenna with a gain of 1 dBi.  Finally, each UE has a RU at a height of 1.5 m and can access the network if the received power from a cell exceeds -110 dBm. 

Given the above network parameters, the macro cell provides coverage up to a 5 km radius, while each micro cell covers a radius of 0.5 km. This implies that up to 75 non-overlapping micro cells can be placed within the coverage area of a single macro cell, enabling the simulation of diverse heterogeneous RAN deployments. Since the micro cells are non-overlapping and maintain LOS connectivity with UEs, inter-cell interference is not considered in our evaluation.

\subsection{UE Mobility and Data Collection Framework}

To simulate realistic UE behavior and collect high-quality training data, we adopt a drive-testing approach where a UE traverses the tracking area and reports signal measurements from all BSs. In our setup, the RSRP is used as the primary indicator of signal strength and channel quality for both handover decisions and model training. The UE measures and reports RSRP values from all BSs at a fixed sampling interval of 50~ms while following a Manhattan mobility model at a constant speed of 60~km/h, over a total simulation duration of 24 hours. Handover decisions are triggered based on thresholds defined by HOM = 0~dB and TTT = 500~ms. During each measurement interval, the UE reports the received RSRP values to the serving BS. The serving BS then forwards both the collected RSRP vector  and the active serving BS label to the Near-RT RIC, where they are stored in the internal database for downstream processing by our ML predictive handover model. Note that this data collection process is fully configurable, enabling variations in the number of BSs, handover parameters, sampling intervals, and reporting schemes. Through this procedure, we construct a 10~GB imbalanced dataset comprising temporal RSRP measurements and corresponding serving BS labels, thereby offering the flexibility required to rigorously evaluate our model under diverse RAN configurations.

\subsection{Model Evaluation and Baselines}
Given that the collected dataset exhibits class imbalance, where certain cells are more frequently used as serving BSs, we report precision, recall, and F1 score in addition to accuracy to enable a fair and robust comparison with state-of-the-art predictive handover models. To ensure consistency, we adopt the same model architecture and training configurations across all baselines, using three fully connected layers with 128 neurons each and ReLU as activation functions. All models were evaluated using an 80/20 train-test split, and 10-fold cross-validation was employed to ensure statistical robustness. More specifically, we benchmark our model against four representative predictive handover algorithms from the literature, each employing a distinct learning architecture:

\noindent \textit{1) LSTM-based Predictive Handover Algorithm}: This approach uses stacked LSTM layers to capture temporal dependencies in past signal measurements and forecast the next serving BS~\cite{b8}.

\noindent \textit{2) GRU-based Predictive Handover Algorithm}: This baseline utilizes Gated Recurrent Units (GRUs) to model sequential input data, offering a lightweight alternative to LSTMs with comparable learning capacity~\cite{b10}.

\noindent \textit{3) RNN-based Predictive Handover Algorithm}: A deep Recurrent Neural Network (RNN) is used to predict upcoming BS associations, as introduced in~\cite{b9}.

\noindent \textit{4) MLP-based Predictive Handover Algorithm}: This method applies a feedforward Multi-Layer Perceptron (MLP) that leverages features extracted from the last \( K \) signal measurements for BS selection~\cite{b6,b7}.


\subsection{Predictive Performance Evaluation}

We analyze the performance of our predictive handover algorithm compared to the described baselines across multiple RAN configurations and temporal windows. Fig.~\ref{fig:model-metrics} presents the results for accuracy, precision, recall, and F1 score across various combinations of the number of BSs \(N\), historical window size \(K\), and prediction window size \(W\).

Across all metrics, our model consistently outperforms the LSTM, GRU, RNN, and MLP baselines, demonstrating improved predictive accuracy and robustness. As shown in Fig.~\ref{fig:model-metrics}(a), accuracy improves with larger temporal input windows \(K\) and longer prediction horizons \(W\). For instance, when moving from \( (K, W) = (10,10) \) to \( (60,60) \), our model maintains an accuracy above 95\%, even as the number of BSs increases from 21 to 51, highlighting strong generalization across spatial and temporal scales. Notably, baseline methods exhibit limited scalability or even performance degradation under increased complexity, particularly in scenarios with 51 BSs and larger window sizes. Similarly, Fig.~\ref{fig:model-metrics}(b–d) shows that our model achieves consistently higher precision, recall, and F1 scores across all settings. This indicates its ability to avoid both false positives and false negatives, which is especially critical in imbalanced datasets where certain BSs dominate the label distribution. The improvement in F1 score becomes more pronounced with larger \(K\) and \(W\), suggesting that our model effectively leverages longer temporal context without overfitting or loss of generality. 

\begin{figure}[t]
    \centering
    \includegraphics[width=0.95\linewidth]{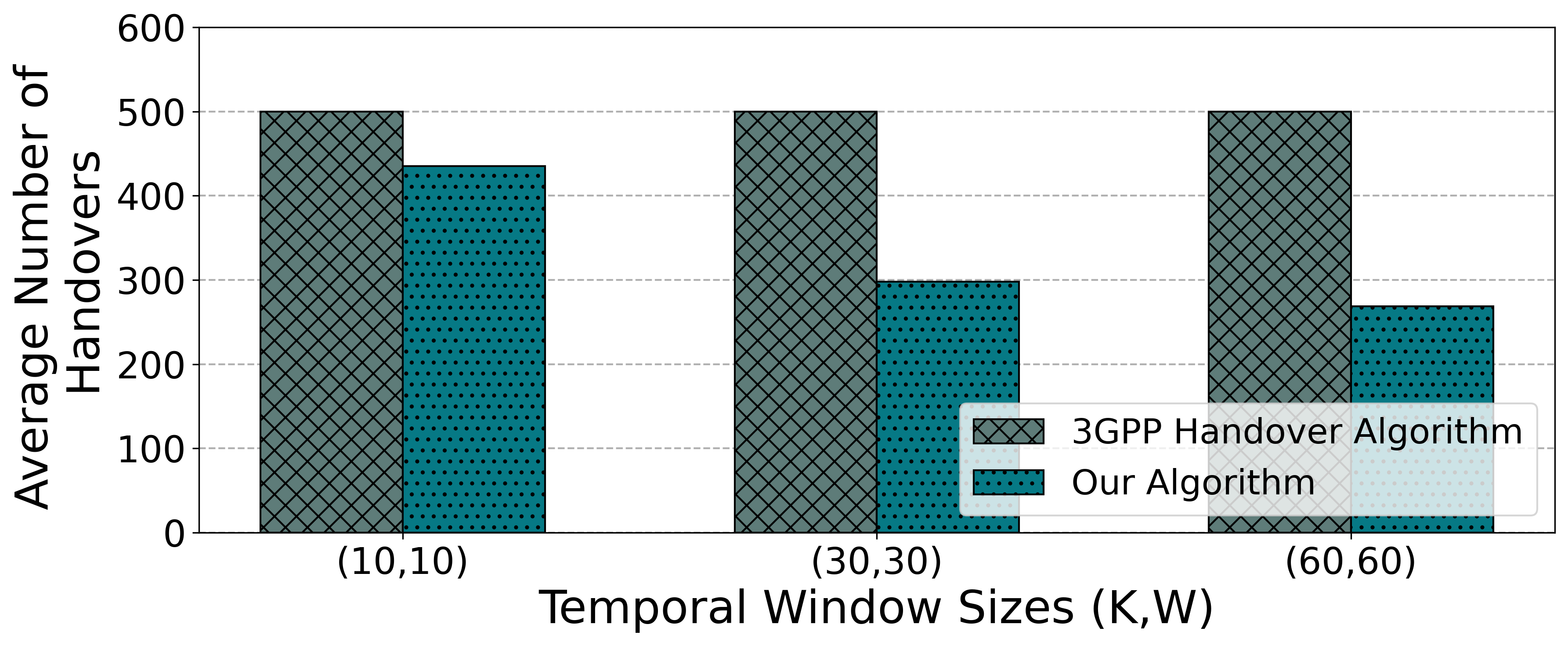}
    \caption{Reduction in handovers achieved by our method compared to the standard 3GPP algorithm across different temporal window configurations \((K, W)\).}
    \label{fig:handover_comparison}
\end{figure}

We further evaluate how our predictive model influences the frequency of handover events. Fig.~\ref{fig:handover_comparison} compares the average number of handovers generated by our model against the standard 3GPP algorithm (explained in section II.B), evaluated over multiple mobility trials. Our approach consistently reduces handover frequency across all window sizes, with reductions ranging from 13\% to 46.2\%. This highlights the practical benefit of our algorithm in minimizing unnecessary handovers while maintaining robust connectivity.

\subsection{Learning in Dynamic RAN Topologies}
\begin{figure}[h!]
    \centering
    \includegraphics[width=\columnwidth]{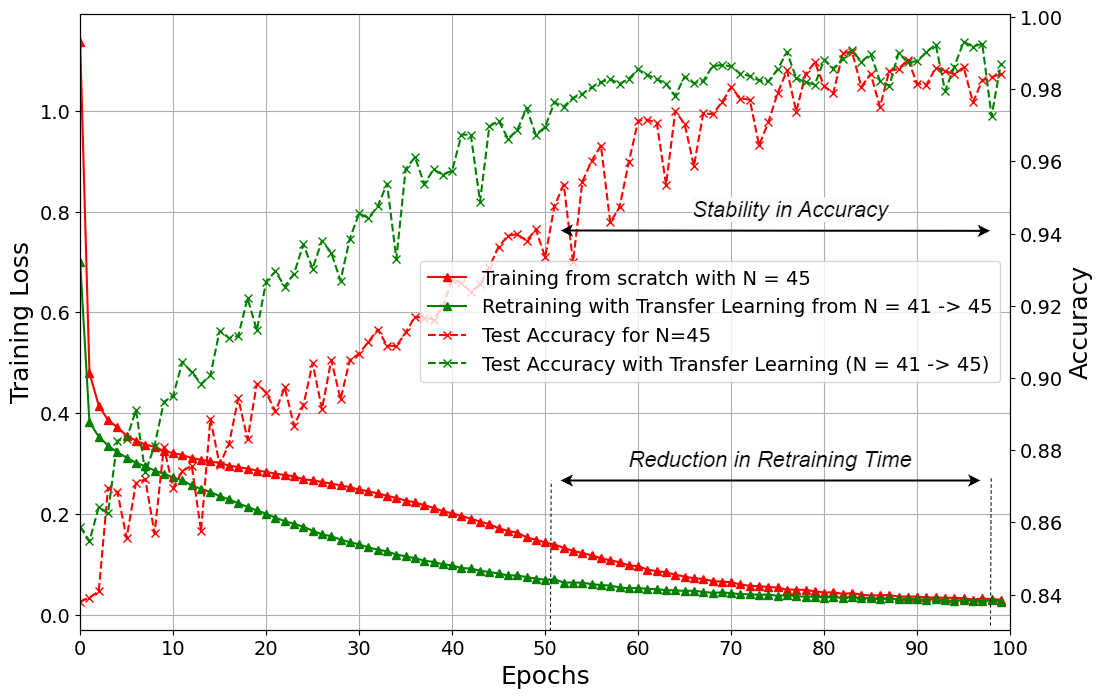} 
    \caption{Reduction in our model's retraining time.}
    \label{fig:transfer_learning} 
\end{figure}

Finally, we assess the adaptability of our model to dynamic RAN topologies by showcasing a scenario in which the number of BSs controlled by the RIC increases from \(N\) to \(N+M\) during inference. To facilitate fast retraining of our model, we utilize transfer learning by applying the previous weights to the new model, allowing for quicker adaptation to evolving network conditions while preserving previously learned knowledge, as described in Algorithm~\ref{alg:dynamic_learning}. We demonstrate this with an example illustrated in Fig.~\ref{fig:transfer_learning}, where we initially assume that our model has been deployed for inference in the Near-RT RIC for a RAN with \(N = 41\) BSs. At a random moment, new \(M = 4\) UAV BSs are introduced at random locations within the macro cell. As shown in Fig.~\ref{fig:transfer_learning}, our adaptable architecture reduces the retraining time by approximately 50\% while maintaining high accuracy (green line) compared to the alternative approach of training from scratch (red line, \(N = 45\)). Finally, Table~\ref{tab:retraining_reduction} shows the retraining time improvements achieved by our architecture across various initial network configurations and dynamic additions of UAV BSs in the RAN. Evidently, our adaptable architecture reduces retraining time by at least 30\% in all cases.

\begin{algorithm}[t]
\small  
\caption{Fast Retraining Through Transfer Learning}\label{alg:dynamic_learning}
\textbf{Input:} 
Pre-trained model parameters \(\theta_{\text{pre-trained}}\), initial number of BSs \(N\), number of added BSs \(M\) after the dynamic event, first Encoder's hidden layer \(h_{l1}\), last S-LSTM Component's hidden state \(h_{\text{lstm}}\).

\textbf{Output:} Updated parameters \(\theta\) of the model.
\begin{algorithmic}[1]

\State \textnormal{Load the pre-trained model from the xApp catalog.}
\State \textnormal{Set model parameters} \(\theta \gets \theta_{\text{pre-trained}}\)
\State \textnormal{Deploy the model in the Near-RT RIC for inference.}
\State \textnormal{Monitor the number of BSs in the system.}

\If{dynamic event occurs}
    \State \textnormal{Update the Encoder's input layer:}
    \State \quad \(model.Enc1 \gets \text{Linear}(N+M, h_{l1})\) 
    \State \textnormal{Update the Decoder's output layer:}
    \State \quad \(model.Dec1 \gets \text{Linear}(h_{\text{lstm}}, N+M)\) 

    \State \textnormal{Randomize weights for newly added BSs.}
    \State \textnormal{Transfer previous S-LSTM weights to the new model.}
\EndIf

\State \textnormal{Retrain the updated model using Algorithm 1.} 

\end{algorithmic}
\end{algorithm}

\begin{table}[t]
\centering
\renewcommand{\arraystretch}{1.5}
\resizebox{\columnwidth}{!}{ 
\begin{tabular}{|c|c|c|c|c|c|c|c|c|c|}
\hline
\textbf{N} & 21 & 21 & 21 & 31 & 31 & 31 & 41 & 41 & 41 \\ 
\hline
\textbf{M} & 4  & 8  & 10 & 4  & 8  & 10 & 4 & 8 & 10 \\ 
\hline
\textbf{Reduction in Retraining Time} & 52\% & 36\% & 30\% & 47\% & 43\% & 40\% & 48\% & 42\% & 40\%\\ 
\hline
\end{tabular}
}
\caption{Reduction in retraining time across different initial numbers of BSs \(N\) and dynamically inserted BSs \(M\).}
\label{tab:retraining_reduction}
\end{table}

\section{Conclusions}
In this paper, we proposed a scalable and adaptable architecture for predictive handover management in O-RAN networks, leveraging temporal UE channel quality measurements. By formulating the task as a multi-class classification problem, our model accurately predicted future BS associations while maintaining flexibility in dynamic RAN environments. Extensive evaluation demonstrated over 95\% accuracy and a significant reduction in handovers, all while minimizing retraining overhead—even under rapidly changing network conditions.


\begin{thebibliography}{00}

\bibitem{b1} W. Saad, M. Bennis and M. Chen, "A Vision of 6G Wireless Systems: Applications, Trends, Technologies, and Open Research Problems," in IEEE Network, vol. 34, no. 3, pp. 134-142, 2020.

\bibitem{b2} M. Polese, L. Bonati, S. D’Oro, S. Basagni and T. Melodia, "Understanding O-RAN: Architecture, Interfaces, Algorithms, Security, and Research Challenges," in IEEE Communications Surveys \& Tutorials, vol. 25, no. 2, pp. 1376-1411, 2023.

\bibitem{b3} M. E. Morocho-Cayamcela, H. Lee and W. Lim, "Machine Learning for 5G/B5G Mobile and Wireless Communications: Potential, Limitations, and Future Directions," in IEEE Access, vol. 7, pp. 137184-137206, 2019.

\bibitem{b4} M. S. Mollel et al., "A Survey of Machine Learning Applications to Handover Management in 5G and Beyond," in IEEE Access, vol. 9, pp. 45770-45802, 2021.

\bibitem{b5} O-RAN Alliance, "O-RAN Use Cases and Deployment Scenarios: Towards Open and Smart RAN," White Paper, 2020.

\bibitem{b6} Z. -H. Huang, Y. -L. Hsu, P. -K. Chang and M. -J. Tsai, "Efficient Handover Algorithm in 5G Networks using Deep Learning," in IEEE Global Communications Conference, Taipei, Taiwan, pp. 1-6, 2020.

\bibitem{b7} C. Lee, H. Cho, S. Song and J. -M. Chung, "Prediction-Based Conditional Handover for 5G mm-Wave Networks: A Deep-Learning Approach," in IEEE Vehicular Technology Magazine, vol. 15, no. 1, pp. 54-62, 2020.

\bibitem{b8} N. Uniyal et al., "Intelligent Mobile Handover Prediction for Zero Downtime Edge Application Mobility," in IEEE Global Communications Conference (GLOBECOM), Madrid, Spain, pp. 1-6, 2021.


\bibitem{b9} A. Masri, T. Veijalainen, H. Martikainen, S. Mwanje, J. Ali-Tolppa and M. Kajó, "Machine-Learning-Based Predictive Handover," in IFIP/IEEE International Symposium on Integrated Network Management (IM), Bordeaux, France, pp. 648-652, 2021.

\bibitem{b10} B. Shubyn, N. Lutsiv, O. Syrotynskyi and R. Kolodii, "Deep Learning based Adaptive Handover Optimization for Ultra-Dense 5G Mobile Networks," in IEEE International Conference on Advanced Trends in Radioelectronics, Telecommunications and Computer Engineering (TCSET), Lviv-Slavske, Ukraine, 2020, pp. 869-872


\bibitem{b11} O. Orhan, V. N. Swamy, T. Tetzlaff, M. Nassar, H. Nikopour and S.Talwar, ”Connection Management xAPP for O-RAN RIC: A Graph
Neural Network and Reinforcement Learning Approach,” in 
IEEE International Conference on Machine Learning and Applications
(ICMLA), Pasadena, CA, USA, pp. 936-941, 2021.


\bibitem{b12} F. Rinaldi et al., "Non-Terrestrial Networks in 5G \& Beyond: A Survey," in IEEE Access, vol. 8, pp. 165178-165200, 2020

\bibitem{b13} S. Alraih, R. Nordin, A. Abu-Samah, I. Shayea and N. F. Abdullah, "A Survey on Handover Optimization in Beyond 5G Mobile Networks: Challenges and Solutions," in IEEE Access, vol. 11, pp. 59317-59345, 2023.

\bibitem{b14} ETSI, "ETSI TR 138 901 V15.0.0 (2018-07): 5G; Study on channel model for frequencies from 0.5 to 100 GHz (3GPP TR 38.901 version 15.0.0 Release 15)," Technical Report, July 2018.

\bibitem{b15} ETSIA, “Procedures for the 5G System,” European Telecommunications Standards Institute (ETSI), Technical Specification (TS) 123 502, June 2018, version 15.2.0 Release 15

\bibitem{b16} O-RAN Alliance, “O-RAN Architecture Description,” O-RAN.WG1.O-RAN-Architecture-Description, Version 5.0, July 2023. 


\end{thebibliography}
\end{document}